\documentstyle[12pt]{article}
\font\titlefont=cmbx10 scaled \magstep4

\def\lprox{\mathrel{\raise .3ex\hbox{$<$\kern-
.75em\lower1ex\hbox{$\sim$}}}}

\def\gprox{\mathrel{\raise .3ex\hbox{$>$\kern-
.75em\lower1ex\hbox{$\sim$}}}}

\font\cm=cmbsy9 at 12pt
\font\cmb=cmbx10 at 12pt
\begin{document}
\input{epsf}

\begin{flushright}
\vspace*{-2cm}
TUTP-97-06 \\ Feb. 24, 1997
\vspace*{2cm}
\end{flushright}

\begin{center}
{\titlefont  A Superluminal Subway: }\\
\vskip .15in
{\titlefont The Krasnikov Tube} \\
\vskip .3in
Allen E. Everett\footnote{email: everett@cosmos2.phy.tufts.edu} and 
Thomas A. Roman\footnote{Permanent address: Department of 
Physics and Earth Sciences,\\  
\indent Central Connecticut State University, New Britain, 
CT 06050 \\
\indent email: roman@ccsu.ctstateu.edu} \\
\vskip .05in
Institute of Cosmology\\
Department of Physics and Astronomy\\
Tufts University\\
Medford, Massachusetts 02155\\
\end{center}
\vskip .05in
\begin{abstract}
The ``warp drive'' metric recently presented by Alcubierre has the 
problem that an observer at the center of the warp bubble is 
causally separated from the outer edge of the bubble wall. Hence 
such an observer can neither create a warp bubble on demand 
nor control one once it has been created. In addition, such a 
bubble requires negative energy densities. One might hope that 
elimination of the first problem might ameliorate the second as well. 
We analyze and generalize a metric, originally proposed by Krasnikov 
for two spacetime dimensions, which does not suffer from the first 
difficulty. As a consequence, the Krasnikov metric has the interesting 
property that although the time for a one-way trip to a distant star 
cannot be shortened, the time for a {\it round trip}, as measured by 
clocks on Earth, can be made arbitrarily short. In our four dimensional 
extension of this metric, a ``tube'' is constructed along the path of an 
outbound spaceship, which connects the Earth and the star. Inside the 
tube spacetime is flat, but the light cones are opened out so as to allow 
superluminal travel in one direction. We show that, although a single 
Krasnikov tube does not involve closed timelike curves, a time machine 
can be constructed with a system of two non-overlapping tubes. 
Furthermore, it is demonstrated that Krasnikov tubes, like warp bubbles 
and traversable wormholes, also involve  unphysically thin layers of 
negative energy density, as well as large total negative energies, 
and therefore probably cannot be realized in practice.
\end{abstract}
\newpage
\baselineskip=13pt

\section{Introduction}
\label{sec:intro}
Alcubierre \cite{A} showed recently, with a specific example, that it is 
possible within the framework of general relativity to warp spacetime 
in a small ``bubblelike'' region in such a way that a spaceship within 
the bubble may move with arbitrarily large speed relative to nearby 
observers in flat spacetime outside the bubble. His model involves 
a spacetime with metric given by (in units where $G=c=1$):
\begin{equation}
ds^2 = -dt^2 \,(1-v^2 f^2(r_0)) - 2 v f(r_0) \, dx\,dt + dx^2 + dy^2 + dz^2 \,,
\label{eq:AMETRIC}
\end{equation}
where
\begin{equation}
r_0 = {[{(x-x_0(t))}^2 + y^2 + z^2]}^{1/2} \,,
\end{equation}
and 
\begin{equation}
v = \frac{dx_0}{dt} \,. \label{eq:v}
\end{equation}
The function $f$ satisfies $f \approx 0$ for $r_0 > R + \delta R$, and 
$f \approx 1$ for $r_0 < R - \delta R$, where $R$ is the bubble radius 
and $\delta R$ is the half thickness of the bubble wall. A suitable form 
for $f$ is given in Ref.\cite{A}. In the limit $\delta R \rightarrow 0$, $f$ 
becomes a step function. Spacetime is then flat outside a spherical 
bubble of radius $R$ centered on the point $\vec r_0 = (x_0(t), 0, 0)$ 
moving with speed $v$ along the $x$-axis, as measured by observers 
at rest outside the bubble. Here $v$ is an arbitrary function of time which 
need not satisfy $v<1$, so that the bubble may attain arbitrarily large 
superluminal speeds as seen by external observers. Space is also flat 
in the region within the bubble where $f=1$, since it follows from 
Eqs.~(\ref{eq:AMETRIC}) and ~(\ref{eq:v}) that, for $f=1$, a locally inertial 
coordinate system is obtained by the simple transformation 
$x' = x- x_0(t)$. Hence an object moving along with the center of the bubble, 
whose trajectory is given by $x'=0$, is in free fall.

As pointed out in Ref.\cite{A}, there are questions as to whether the metric 
~(\ref{eq:AMETRIC}) is physically realizable, since the corresponding 
energy-momentum tensor, related to it by the Einstein equation, involves 
regions of negative energy density, i. e., violates the weak energy condition 
(WEC) \cite{HE}. This is not surprising since it has been shown \cite{AE} 
that a straightforward extension of the metric of Ref.\cite{A} leads to a 
spacetime with closed timelike curves (CTCs). It is well-known that 
negative energy densities are required for the existence of stable 
Lorentzian wormholes \cite{MTY}, 
where CTCs may also occur, and Hawking \cite{H} has shown that the 
occurrence of CTCs requires violations of the WEC under rather general 
circumstances. The occurrence of regions with negative energy density 
is allowed in quantum field theory \cite{EGJ,Kuo}. However, 
Ford and Roman \cite{FR95,FRWH,FR97} have proven inequalities which 
limit the magnitude and duration of negative energy density. 
These ``quantum inequalities'' (QIs) strongly suggest \cite{FRWH} 
that it is unlikely that stable Lorentzian wormholes can exist, and 
similar conclusions have been drawn by Pfenning and Ford \cite{PFWD} 
with regard to the ``warp drive'' spacetime of Ref.\cite{A}.

It is the goal of this paper to analyze a spacetime recently proposed by 
Krasnikov \cite{K} which, although differing from that of Ref.\cite{A} in 
several key respects, shares with it the property of allowing 
superluminal travel. We first review this spacetime, 
in the two dimensional form as originally given by Krasnikov, 
and give a more extended discussion of its properties than 
that provided in Ref.\cite{K}. Then we carry out the straightforward 
task of generalizing the Krasnikov metric to the realistic case of 
four dimensions. We establish 
that, despite their differences, the Krasnikov and Alcubierre metrics 
share a number of important properties. In particular, we show that, 
like the metric of Ref.\cite{A}, the Krasnikov metric implies the existence 
of CTCs, and also show explicitly that the associated energy-momentum 
tensor violates the WEC. Finally, we apply a QI to the Krasnikov spacetime 
and argue that, as in the cases of wormholes and Alcubierre bubbles, the 
QI strongly suggests that the Krasnikov spacetime is not physically 
realizable.

\section{The Krasnikov Metric in Two Dimensions and Superluminal Travel}
\label{sec:II}

Krasnikov \cite{K} raised an interesting problem with methods of 
superluminal travel similar to the Alcubierre mechanism. The basic 
point is that in a universe described by the Minkowski metric at $t=0$, an 
observer at the origin, e. g., the captain of a spaceship, can do nothing to 
alter the metric at points outside the usual future light cone 
$|\vec r| \leq t$, 
where $r= {(x^2 + y^2 + z^2)}^{1/2}$. In particular, this means that those on 
the spaceship can never create, on demand, an Alcubierre bubble 
with $v > c$ around the ship. This follows explicitly from the following simple 
argument. Points on the outside front edge of the bubble are always 
spacelike separated from the center of the bubble. One can easily 
show this by considering the trajectory of a photon emitted in the positive 
$x$-direction from the spaceship. If the spaceship is at rest at 
the center of the bubble, then initially the photon has 
$dx/dt = v + 1$ or $dx'/dt = 1$. This 
of course must be true since at the center of the bubble the primed 
coordinates define a locally inertial reference frame. However, at 
some point with $x'=x'_c$, for which $0 < f < 1$ so that $x'_c < R$ 
and the point is within the bubble wall, one finds that $dx'/dt = 0$ or 
$dx/dt = v$. (It is clear by continuity that $dx/dt = v$ at some point for 
photons moving in the $+x$-direction inside the bubble, since 
$dx/dt = v+1$ at the center of the bubble and $dx/dt = 1$ in flat space 
outside the bubble wall.) Thus once photons reach $x'_c$, they 
remain at rest relative to the bubble and are simply carried along with it. 
Photons emitted in the forward direction by the spaceship never reach 
the outside edge of the bubble wall, which therefore lies outside the 
forward light cone of the spaceship. The bubble thus cannot be 
created (or controlled) by any action of the spaceship crew, excluding 
the use of tachyonic signals \cite{PUN}.

The foregoing discussion does not mean that Alcubierre bubbles, if it were 
possible to create them, could not be used as a means of superluminal 
travel. It only means that the actions required to change the metric and 
create the bubble must be taken beforehand by some observer whose 
forward light cone contains the entire trajectory of the bubble. 
Suppose space has been warped to create a bubble travelling from the 
Earth to some distant star, e. g., Deneb, at superluminal speed. 
A spaceship appropriately located with respect to the bubble trajectory 
could then choose to enter the bubble, rather like a passenger catching 
a passing trolley car, and thus make the superluminal journey. However, 
a spaceship captain hoping to make use of a region of spacetime with a 
suitably warped metric to reach a star at a distance $D$ in a time interval 
$\Delta t <  D$ must, like the potential trolley car passenger, hope that 
others have previously taken action to provide a passing mode of 
transportation when desired.

In contrast, as Krasnikov points out, causality considerations do not 
prevent the crew of a spaceship from arranging, by their own actions, 
to complete a {\it round trip} from the Earth to a distant star and back 
in an arbitrarily short time, as measured by clocks on the Earth, by 
altering the metric along the path of their outbound trip. As an example, 
consider the metric in the two dimensional $t,x$ subspace, introduced 
by Krasnikov in Ref.\cite{K}, given by 
\begin{eqnarray}
ds^2 &=& - (dt - dx) \, (dt + k(x,t) \, dx)  \label{eq:K2D0} \\ 
&=& -dt^2 + (1- k(x,t)) \, dx\,dt + k(x,t)\, dx^2 \,,
\label{eq:K2D}
\end{eqnarray}  
where 
\begin{equation}
k(x,t) \equiv 1- (2 - \delta) \, \theta_{\epsilon} (t - x) \, 
[ \theta_{\epsilon} (x) - \theta_{\epsilon}(x + \epsilon - D) ] \,,
\label{eq:KDEF}
\end{equation}
$\theta_{\epsilon}$ is a smooth monotonic function satisfying 
\begin{equation}
\theta_{\epsilon}(\xi)  = \left\{\matrix{1 & {\rm at} \,\, \xi > \epsilon\cr
0& {\rm at} \,\, \xi < 0}\right. \,,
\label{eq:DEFTHETA}
\end{equation}
and $\delta$ and $\epsilon$ are arbitrary small positive parameters. 
We will give a specific form for $\theta_\epsilon$ in Sec.~\ref{sec:QI}. 
For $k=1$, the metric~(\ref{eq:K2D}) reduces to the two dimensional 
Minkowski metric. The two time-independent $\theta_\epsilon$-functions 
between the brackets in Eq.~(\ref{eq:KDEF}) vanish for $x < 0$ and 
cancel for $x > D$, ensuring $k=1$ for all $t$ except between $x=0$ and 
$x=D$. When this behavior is combined with the effect of the factor 
$\theta_\epsilon(t-x)$, one sees that the metric ~(\ref{eq:K2D}) describes 
Minkowski space everywhere for $t<0$, and at all times outside the range 
$0<x<D$. For $t>x$ and $\epsilon < x < D - \epsilon$, the first two 
$\theta_\epsilon$-functions in Eq.~(\ref{eq:KDEF}) both equal $1$, 
while $\theta_\epsilon(x + \epsilon - D) = 0$, giving $k = \delta -1$ 
everywhere within this region. There are two spatial boundaries 
between these two regions of constant $k$, one between $x=0$ 
and $x = \epsilon$ for $t>0$ and a second between $x=D- \epsilon$ 
and $x = D$ for $t>D$. We can think of this metric as being produced 
by the crew of a spaceship which departs from Earth ($x=0$) at $t=0$ and 
travels along the $x$-axis to Deneb ($x=D$) at a speed which for simplicity 
we take to be only infinitesimally less than the speed of light, so that it 
arrives at $t \approx D$. The crew modify the metric by changing $k$ 
from $1$ to $\delta - 1$ along the $x$-axis in the region between $x=0$ and 
$x=D$, leaving a transition region of width $\epsilon$ at each end to insure 
continuity. Similarly, continuity in time implies that the modification of $k$ 
requires a finite time interval whose duration we assume, again for 
simplicity, to be $\epsilon$. However, since the boundary of the forward 
light cone of the spaceship at $t=0$ is given by $|x| = t$, the spaceship 
cannot modify the metric at $x$ before $t=x$, which accounts for the 
factor $\theta_\epsilon(t-x)$ in the metric. Thus there is a transition region 
in time between the two values of $k$, of duration $\epsilon$, lying along 
the world line of the spaceship, $x \approx t$. The resulting geometry in 
the $x-t$ plane is shown in Fig. 1, where the shaded regions represent the 
two spatial transition regions $0< x < \epsilon$ and $D - \epsilon < x < D$ 
and the temporal transition region $x < t  <  x + \epsilon$. In the internal 
region of the diagram, enclosed by the three shaded areas, $k$ has the 
constant value $\delta - 1$, while $k=1$ everywhere outside the shaded 
regions. The world line of the spaceship is represented by the line $AB$.
\begin{figure}
\begin{center}\leavevmode\epsfysize=15cm\epsffile{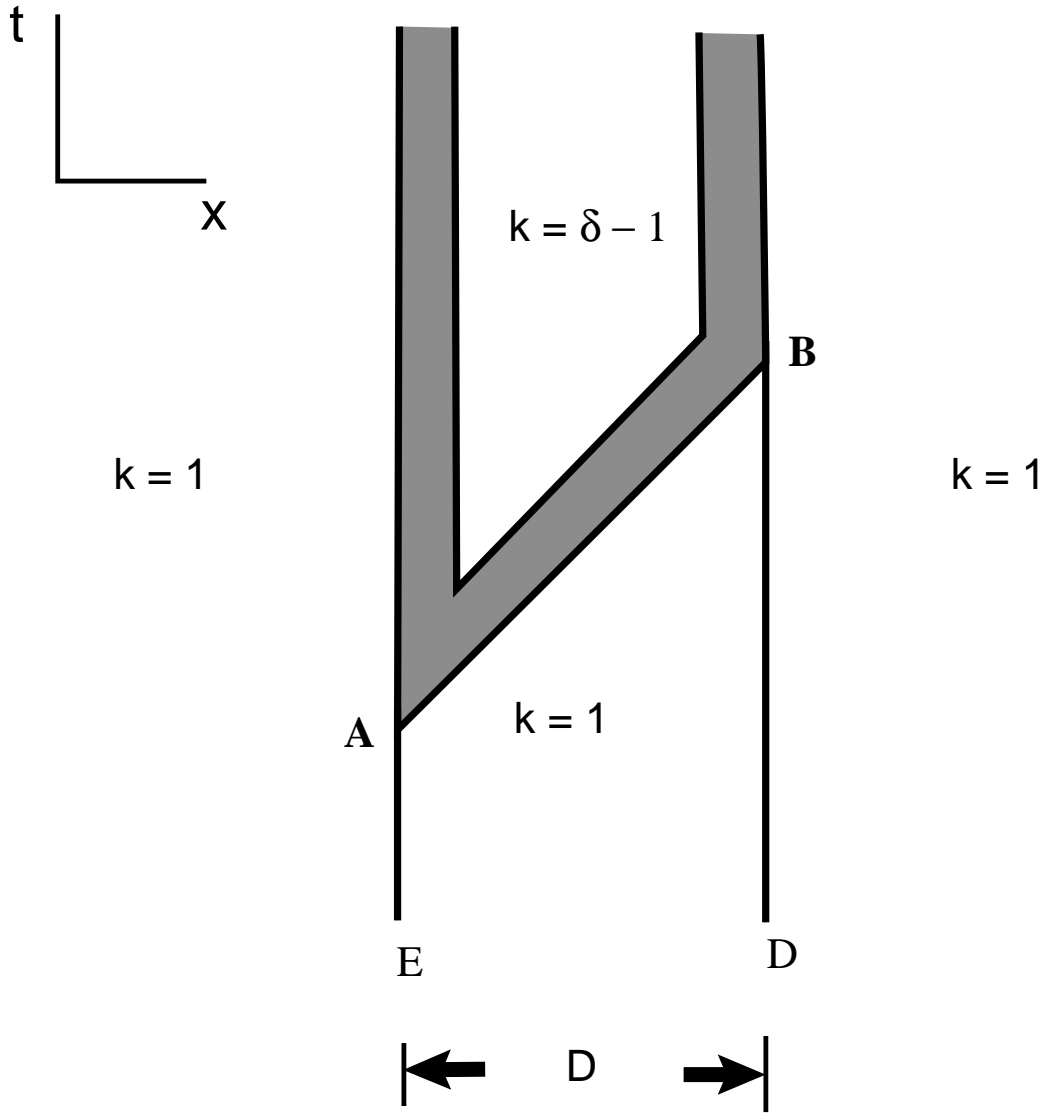}\end{center}
\caption{The Krasnikov spacetime in the $x-t$ plane. The vertical lines 
{\cm E} and {\cm D} are the world lines of the Earth and 
Deneb, respectively. The world line of the spaceship is (approximately) 
represented by the line {\cmb AB}.}
\end{figure}

The properties of the modified metric with $\delta-1 \leq k \leq 1$ can be 
easily seen from the factored form of the expression for $ds^2$ in 
Eq.~(\ref{eq:K2D0}) where, putting $ds^2 = 0$, one sees that the two 
branches of the forward light cone in the $t,x$ plane are given by 
$dt/dx = 1$, and $dt/dx = -k$. As $k$ becomes smaller and then 
negative, the slope of the left-hand branch of the light cone becomes 
less negative and then changes sign; i. e., the light cone along the 
negative $x$-axis ``opens out''. This is illustrated in Fig. 2 
where we depict the behavior of the light cone (in two 
spatial dimensions) for $k=1,\, k=0$, and $k= \delta -1$ for 
small $\delta$. 
\begin{figure}
\begin{center}\leavevmode\epsfysize=5cm\epsffile{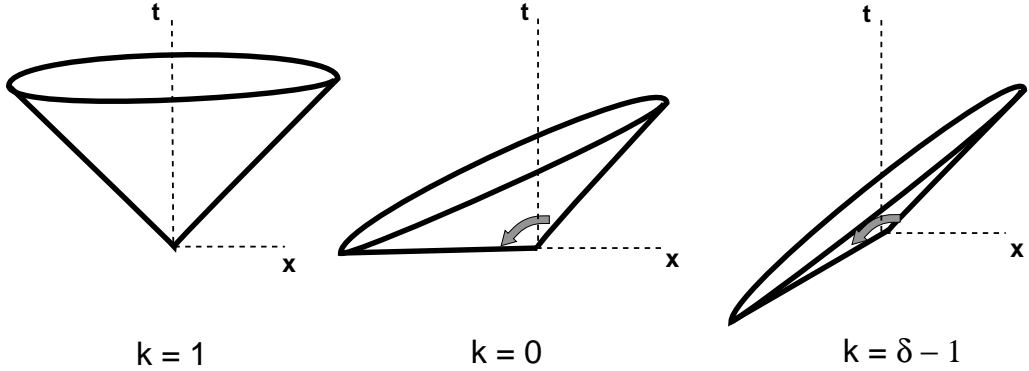}\end{center}
\caption{Forward light cones in the Krasnikov spacetime (illustrated 
with two space dimensions) for $k=1,\,k=0$, and $k=\delta-1$.}
\end{figure}
For $k \approx -1$, the boundary of the forward light cone 
is almost the straight line $x = t$, and the forward and backward light 
cones include almost all of spacetime.

In the internal region of Fig. 1, where $k= \delta -1= const$, 
space is flat, since the metric of Eq.~(\ref{eq:K2D}) can be reduced to 
Minkowski form by the coordinate transformation
\begin{eqnarray}
dt' &=& dt + \left( \frac{\delta}{2} -1 \right)\, dx\,, \label{eq:DT'}   \\
dx' &=& \left( \frac{\delta}{2}\right) \,dx \,.    \label{eq:DX'}
\end{eqnarray}
Note that the left-hand branch of the light cone in the internal region 
is given in the Minkowski coordinates by $dt'/dx' = -1$, which, from 
Eqs.~(\ref{eq:DT'}) and ~(\ref{eq:DX'}), reduces to our previous 
expression $dt/dx = - k = 1- \delta$ on the left-hand branch of the 
light cone as illustrated in Fig. 2. We also note that the 
transformation becomes singular at $\delta = 0$, i. e., at $k= -1$. 

From Eqs.~(\ref{eq:DT'}) and ~(\ref{eq:DX'}), we obtain
\begin{equation}
\frac{dt}{dt'} = 1 + \left( \frac{2 - \delta}{\delta} \right) \, 
\frac{dx'}{dt'} \,. 
\label{eq:DT/DT'}
\end{equation}
For an object propagating causally, i. e., into its forward light cone, we 
have $|dx'/dt'| < 1$ and $dt' > 0$. Since $0 < \delta \leq 2$, one sees that 
for such an object moving in the positive $x'$ (and $x$) direction, 
$dt > 0$ for any $\delta$. However, for $\delta<1$, an object moving 
sufficiently close to the left branch of the light cone given by 
$dx'/dt' = -1$, will have $dt/dt' < 0$ and thus appear to propagate 
backward in time as seen by observers in the external 
($\delta=2\,,k=1$) region of Fig. 1. These properties of motion in 
the Krasnikov metric with $\delta<1$ can be seen from the shape of 
the light cone as shown in Fig. 2.  

Now suppose our spaceship, having travelled from the Earth to Deneb 
and arriving at time $t \approx D$, has modified the metric so that 
$k \approx -1$ (i. e., $\delta \approx 0$) along its path. Suppose further 
that it now returns almost immediately to Earth, again travelling at a 
speed arbitrarily close to the speed of light as seen in its local inertial 
system, i. e., along the left-hand branch of the light cone with 
$dx'/dt' \approx -1$. It will then have $v_r \equiv dx/dt \approx -1/k =
1/(1- \delta) \approx 1$ and $dt<0$ (since $dx<0$), and thus move 
down and to the left along the upper edge of the diagonal shaded 
region in Fig. 1. The spaceship's return to Earth requires a time interval 
$\Delta t_r = - D/v_r = D(\delta -1)$, and the ship returns to Earth at a time 
$t_E$ as measured by clocks on the Earth given by 
$t_E = D + \Delta t_r = D \delta$. (For simplicity, here we treat the wall 
thickness, $\epsilon$, as negligible.)

Since $\delta > 0$, $|\Delta t_r| < D$, and the spacetime interval between 
the spaceship's departure from Deneb and its return to Earth is spacelike. 
Therefore the return journey must involve superluminal travel. Note that 
$t_E > 0$, meaning that the return of the spaceship to Earth 
necessarily occurs after its departure. However, the interval between 
departure and return, as measured by observers on the Earth, can be 
made arbitrarily small by appropriate choice of the parameter $\delta$. 
The time of return, $t_E$ must necessarily be positive, since causality 
insures that the metric is modified, opening out the light cone to allow 
causal propagation in the negative $t$-direction, only for $t>0$. Since 
$t_E>0$, the spaceship cannot travel into its own past; i. e., the 
metric of Ref.\cite{K}, as it stands, does not lead to CTCs and the 
existence of a time machine. However, when the metric ~(\ref{eq:K2D}) 
is generalized to the more realistic case of three space dimensions 
CTCs do become possible, as we shall see below.

Before turning to the three dimensional generalization, we note one 
other interesting property of the Krasnikov metric. 
In the case $\delta<1$, it is always 
possible to choose an allowed value of $dx'/dt'$ for which $dt/dt' = 0$, 
meaning that the return trip is instantaneous as seen by observers in 
the external region of Fig. 1. This can be seen from the third diagram 
in Fig. 2. It also follows easily from Eq.~(\ref{eq:DT/DT'}), which implies 
that $dt/dt' = 0$ when $dx'/dt'$ satisfies 
\begin{equation}
\frac{dx'}{dt'} = - \frac{\delta}{(2- \delta)} \,,
\label{eq:DX'/DT'I}
\end{equation}
which lies between $0$ and $-1$ for $0< \delta < 1$.
 
\section{Generalization to Four Dimensions}
\label{sec:4D}
In four dimensions the modification of the metric begins along the path 
of the spaceship, i. e., the $x$-axis, occurring at position $x$ at time 
$t \approx x$, the time of passage of the spaceship. We assume 
that the disturbance in the metric propagates radially outward from 
the $x$-axis, so that causality guarantees that at time $t$ the region 
in which the metric has been modified cannot extend beyond 
$\rho = t - x$, where $\rho ={(y^2 + z^2)}^{1/2}$. It also seems natural 
to take the modification in the metric not to extend beyond some 
maximum radial distance $\rho_{max} \ll D$ from the $x$-axis. Thus in 
four dimensional spacetime we replace Eq.~(\ref{eq:KDEF}) by 
\begin{equation}
k(t,x,\rho) \equiv 1- (2 - \delta) \, \theta_{\epsilon} (\rho_{max}- \rho) \, 
\theta_{\epsilon} (t - x - \rho) \, 
[ \theta_{\epsilon} (x) - \theta_{\epsilon}(x + \epsilon - D) ] \,,
\label{eq:KDEF4D}
\end{equation}
and our metric, now written in cylindrical coordinates, is given by 
\begin{equation}
ds^2  = -dt^2 + (1- k(t,x,\rho)) \, dx\,dt + k(t,x,\rho)\, dx^2 
+ d{\rho}^2 + {\rho}^2 \, d{\phi}^2 \,.
\label{eq:K4D}
\end{equation}  
(Again we assume for simplicity that the $\epsilon$-parameters in 
all of the $\theta_\epsilon$-functions which appear in the expression 
for $k$ are equal.) For $t \gg D + \rho_{max}$ one now has a tube of 
radius $\rho_{max}$ centered on the $x$-axis within which the metric 
has been modified; we refer to this structure as a ``Krasnikov tube''. 
In contrast with the metric of Ref.\cite{A}, the metric of a Krasnikov 
tube is static once it has been initially created. If we make the 
assumption that $\rho_{max} \gg \epsilon$, such a tube will consist of a 
relatively large central core, of radius $\rho_{max} - \epsilon$, along the 
$x$-axis with $\epsilon < x < D - \epsilon$; within this central core 
space is flat and $k = \delta -1 = const$. This central core will be 
surrounded by thin walls and end caps of thickness $\epsilon$, 
within which there is curved space with $k$ varying between 
$k= \delta -1$ and $k=1$. The situation is illustrated in Fig. 3, which 
shows cross-sections through the tube in the region 
$\epsilon < x < D - \epsilon$ and also in one of the end caps.
\begin{figure}
\begin{center}\leavevmode\epsfysize=7.75cm\epsffile{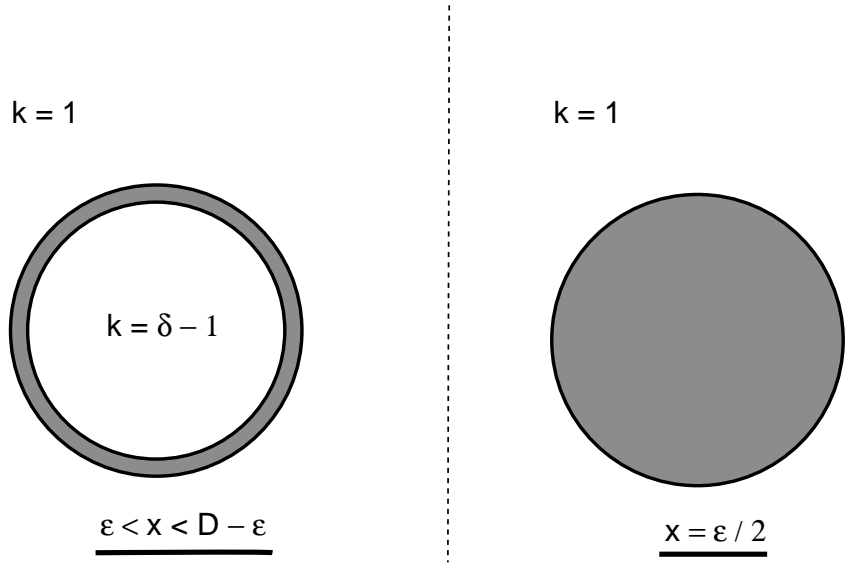}\end{center}
\caption{Spatial cross-sections of a Krasnikov tube at $x=const$, 
$t=const$. The left diagram represents a cross-section through the 
middle of the tube between the end caps, while the right diagram is 
a cross-section through an end cap.} 
\end{figure}

\section{A Superluminal Subway and Closed Timelike Curves}
\label{sec:TM}
As we have seen, in two dimensions a single Krasnikov tube allows 
superluminal travel backward in $t$ in one direction along the $x$-axis, 
and does not lead to CTCs. However, in three space dimensions the 
situation is different. Assuming that Krasnikov tubes can be established, 
imagine that a spaceship has travelled from the Earth to Deneb along 
the $x$-axis during the time interval from $t=0$ to $t=D$, and established 
the Krasnikov tube running from the Earth to Deneb which we have 
discussed. It would then be possible for the ship to return from Deneb 
to the Earth outside the first tube along a path parallel to the $x$-axis 
but at a distance $\rho_0$ from it, where $D \gg \rho_0 > 2 \rho_{max}$. 
On the return journey the spaceship crew could again modify the 
metric along their path, establishing a second Krasnikov tube identical 
to the first but running in the opposite direction; that is, the metric within 
the second tube would be given by that of Eq.~(\ref{eq:K4D}) with 
$x$ replaced by $X \equiv D - x$ and $t$ replaced by $T \equiv t - D$. 
The crucial point is that in three dimensions the two tubes can be made 
non-overlapping because of their separation in the $\rho$-direction. 
The spaceship can now, for example, start from the Earth ($x=0$) 
at $t= 2D$, and travel back in time to the Earth at a time arbitrarily close 
to $t=0$ by first using the second Krasnikov tube to travel to Deneb 
($x=D$) at time $t=D$, and then using the original tube as before, to 
travel from Deneb at $t=D$ to the Earth at $t=0$. (We are assuming 
that the ship travels at essentially light speed, that $\delta$ and 
$\epsilon$ are taken to be negligibly small, 
and that the small time required to move the 
distance $\rho_0 \ll D$ from one of the Krasnikov tubes to the other is 
negligible.) It may be worth noting that the foregoing argument is 
closely analogous to that given in Ref.\cite{AE} for the existence 
of CTCs in the Alcubierre case. The situation is also similar to the 
case of time travel using a two-wormhole system, as depicted in 
the spacetime diagram of Fig. 18.5 of Ref.\cite{VBOOK}. 

It follows from the foregoing discussion that if Krasnikov tubes could be 
constructed, one could, at least in principle, establish a network of such 
tubes forming a sort of interstellar subway system allowing 
instantaneous communication between points connected by the tubes. 
A necessary corollary of the existence of such a network is the possibility 
of backward time travel and the consequent existence of CTCs. CTCs 
could be avoided only if, for some reason, there existed a preferred 
axis such that all the Krasnikov tubes were oriented 
so that the velocity components 
along that axis of objects in superluminal motion were always positive, 
implying that no object could return to the same point in time {\it and} 
space. One might be tempted to reject immediately the possibility of 
Krasnikov tubes for interstellar travel because, unlike Alcubierre bubbles, 
their required length would be enormous. However, there are 
interesting situations in astronomy, e. g., jets in active galactic nuclei 
and possibly cosmic strings, which involve (albeit positive) matter 
distributions of such dimensions. In any event, even if the construction 
of Krasnikov tubes over astronomical distances is impractical, 
oppositely directed non-overlapping pairs of tubes of laboratory 
dimensions could form time machines, forcing one to confront 
all the associated problems. 

\section{The Stress-Energy Tensor for a Krasnikov Tube}
\label{sec:Tmunu}
In this section, we show that the WEC is necessarily violated in some 
regions of a Krasnikov tube. The stress-energy tensor $T_{\mu\nu}$ 
for the matter needed to produce a Krasnikov tube can be calculated 
from the metric of Eq.~(\ref{eq:K4D}) and the Einstein equations, using 
the program MATHTENSOR \cite{PC}. We first obtain an expression 
for $T_{\mu\nu}$ in terms of derivatives of $k$ with respect to the 
spacetime coordinates. The stress tensor element $T_{tt}$ is given 
by the following expression:
\begin{equation}
T_{tt} = {\left[{32 \,\pi \,{(1+k)}^2}\right]}^{-1} \,
\biggl[ { \frac{-4 \, (1+k)}{\rho} \, \frac{\partial k}{\partial \rho} \, + \, 
3 {\biggl( \frac{\partial k}{\partial \rho} \biggr) }^2 \, - 
\, 4\,(1+k)\,  \frac{ {\partial}^2 k}{{\partial \rho}^2} } \biggr]\,.
\label{eq:T}
\end{equation}
(It will be shown later that 
this is the energy density seen by a static observer.) 
Note that this component of the stress tensor involves only 
derivatives of $k$ with respect to $\rho$.  A number of 
general features of the $k$ vs. $\rho$ curve illustrated in Fig. 4 are 
generic and follow from Eq.~(\ref{eq:KDEF4D}) without specifying an 
explicit form for $\theta_\epsilon$. 
\begin{figure}
\begin{center}\leavevmode\epsfysize=15cm\epsffile{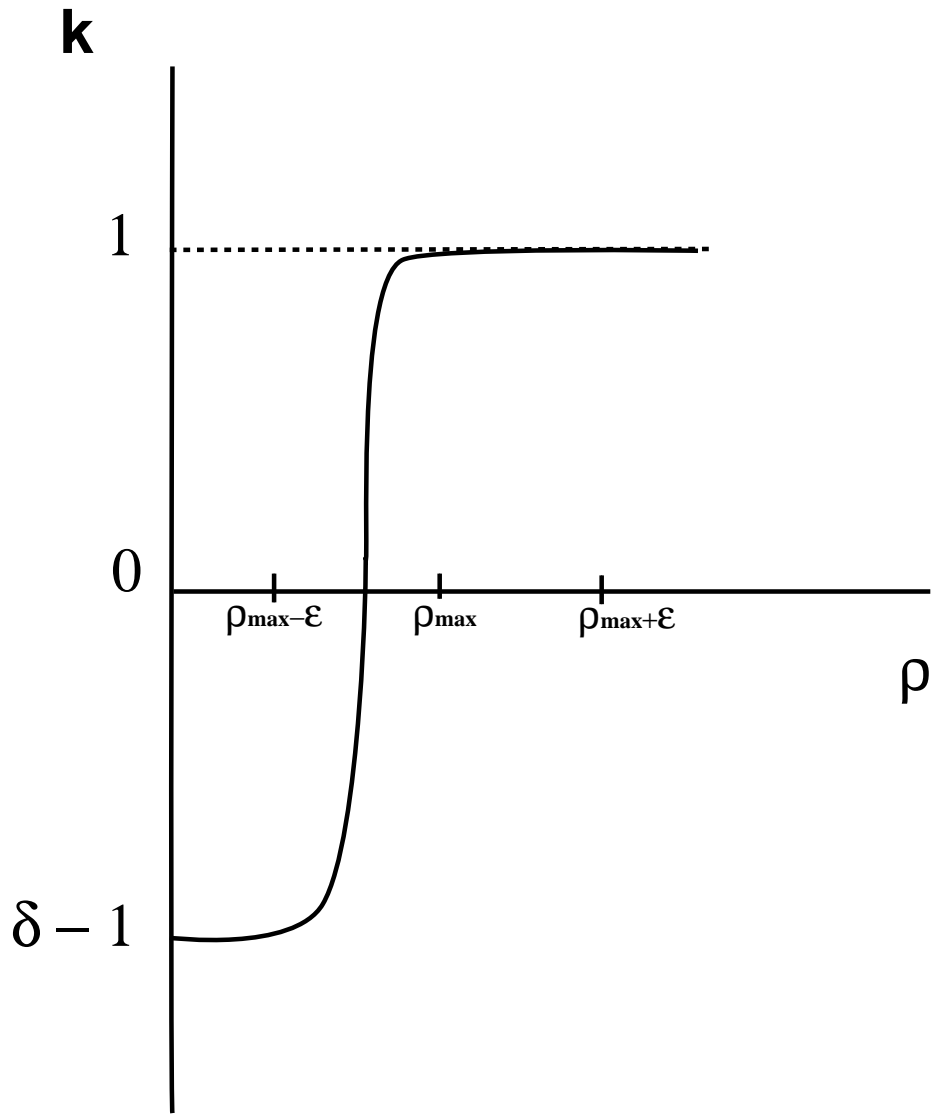}\end{center}
\caption{Graph of $k$ vs $\rho$ at constant $x,t$. 
Here $\epsilon< x < D - \epsilon$.}
\end{figure}
In particular, $k$ increases 
monotonically from its value at $\rho=0$ to $k \approx 1$ at 
$\rho \geq \rho_{max}$, so that $\partial k / \partial \rho$ and 
$(1+k)$ are positive. Furthermore, analyticity of $k$ at $\rho = 0$ 
implies that $\partial k / \partial \rho$ vanishes at that point. From 
the previous remarks, we have that 
$\partial k / \partial \rho \approx \beta {\rho}^m$, with $m \geq 1$, 
$\beta > 0$, for small $\rho$. Hence, sufficiently near the axis of the 
Krasnikov tube, the first and third terms on the right-hand side of 
Eq.~(\ref{eq:T}) are negative and go as ${\rho}^{m-1}$ for 
$\rho \rightarrow 0$; these terms thus dominate the second term, 
which is positive, by a factor of ${\rho}^{-m-1}$. Therefore there is 
necessarily a range of $\rho$ near the axis of the tube where the 
energy density seen by a static observer is negative.

As we have noted previously, for a thin-walled tube, space is nearly 
flat and so $T_{tt} \approx 0$ within the core of the tube. Thus in the 
region near $\rho = 0$ where we can make a general statement 
about the sign of $T_{tt}$ on the basis of the preceding argument, 
we expect $T_{tt}$ to be extremely small due to the behavior of the 
function $\theta_\epsilon(\rho_{max} - \rho)$. (We observe that 
the case $k = -1$, corresponding to $\delta =0$, is not allowed, since 
that would produce a divergence in $T_{tt}$.) In the vicinity of the 
tube wall, where $T_{tt}$ is large, we can only obtain its value by 
choosing an explicit form for $k$, i. e., for $\theta_\epsilon$, and 
then evaluating $T_{tt}$ numerically. In Fig. 5 we show a plot of 
$T_{tt}$ in the region of the tube wall obtained in this way, using 
the form of $\theta_\epsilon$ given in the next section, and taking 
$x = D/2$, $t \gg \rho_{max} + D/2$, and $\epsilon / \rho_{max} \ll 1$. 
\begin{figure}
\begin{center}\leavevmode\epsfysize=7.5cm\epsffile{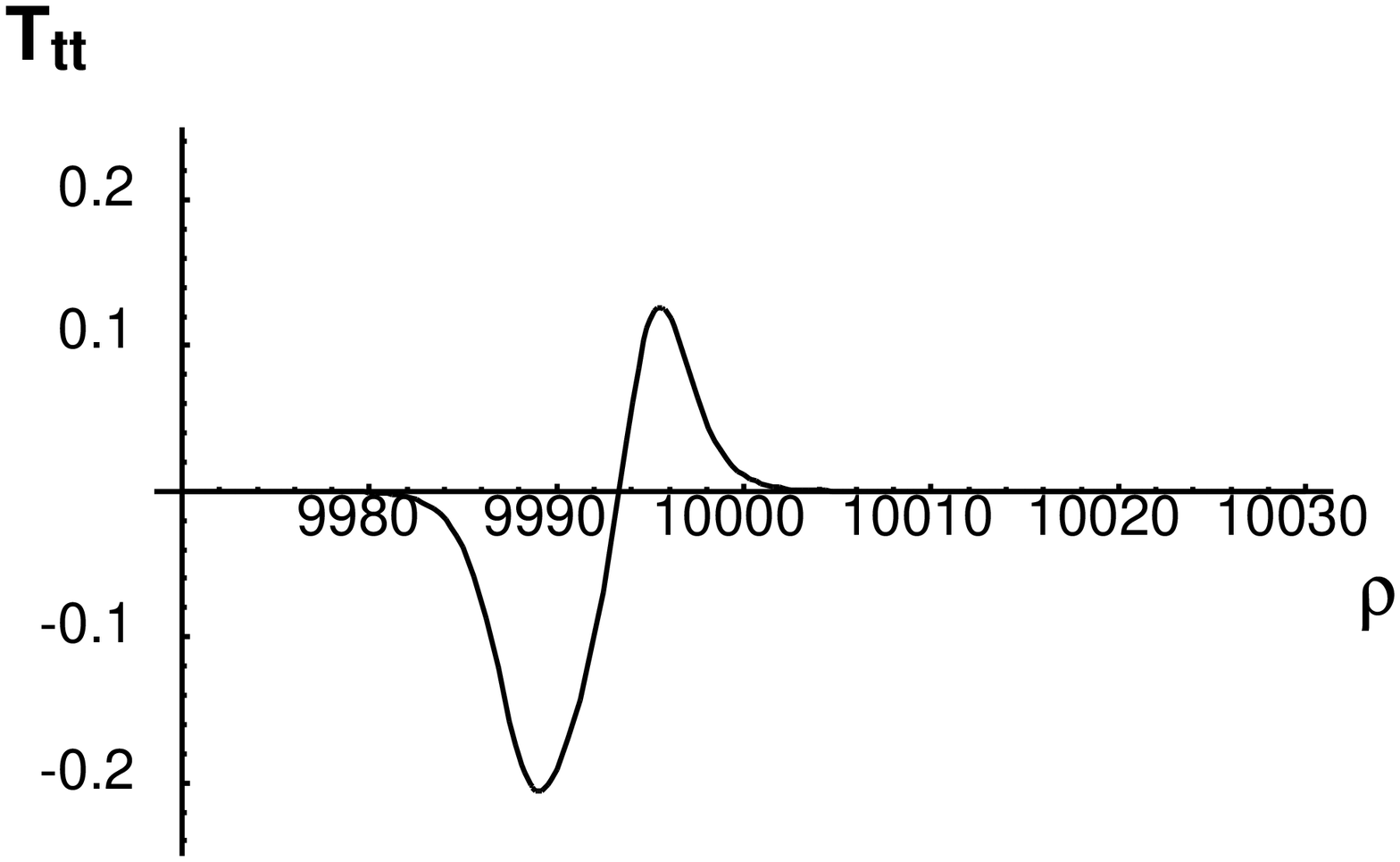}\end{center}
\caption{Graph of energy density vs. $\rho$ at the middle of the 
tube, i. e., at $x = D/2$ and $t= const \gg D/2 + \rho_{max}$. 
Here we have chosen $\delta = 0.01,\,
\epsilon = 10$, and $\rho_{max} = 1000 \, \epsilon$. The plot 
extends from $\rho_{max} - 3 \epsilon$ to $\rho_{max} + 3 \epsilon$.}
\end{figure}
We see that $T_{tt}$ is negative on the inner side of the wall, as one 
would expect, since the general argument given above shows that 
$T_{tt}$ must be negative for small $\rho$. However, $T_{tt}$ changes 
sign and develops appreciable positive values on the outer side of the 
wall. The corresponding plot at $x=\epsilon/2$, in the left endcap, is 
very similar to Fig. 5. There are two main differences: first, the magnitudes 
of the positive and negative energy density maximum and minimum are 
essentially equal; and second, these magnitudes are roughly four times 
smaller than at $x=D/2$.  

\section{Quantum Inequality Constraints}
\label{sec:QI}      
In Ref.\cite{FR95}, an inequality was proven which limits the magnitude
and duration of the negative energy density seen by an inertial observer in
Minkowski spacetime. Let $\langle T_{\mu\nu} \rangle$
be the renormalized expectation value of the stress tensor for a free,
massless, minimally coupled scalar field in
an arbitrary quantum state. Let $u^\mu$ be the observer's four-velocity, so
that $\langle T_{\mu\nu} u^{\mu} u^{\nu}\rangle$ is the expectation value of
the local energy density in this observer's frame of reference. The inequality
states that
\begin{equation}
{{\tau_0} \over \pi}\, \int_{-\infty}^{\infty}\,
{{\langle T_{\mu\nu} u^{\mu} u^{\nu}\rangle\, d\tau}
\over {{\tau}^2+{\tau_0}^2}} \geq
-{3\over {32 {\pi}^2 {\tau_0}^4}}\,,  \label{eq:QI}
\end{equation}
for all $\tau_0$, where $\tau$ is the observer's proper time.
The Lorentzian function which appears in the integrand is a convenient 
choice for a sampling function, which samples the energy density in 
an interval of characteristic duration $\tau_0$ centered around an 
arbitrary point on the observer's worldline. The proper time coordinate 
has been chosen so that this point is at $\tau = 0$. Physically, 
Eq.~(\ref{eq:QI}) implies that the more negative the
energy density is in an interval, the shorter must be the duration of the
interval.  Such a bound is called a  ``quantum inequality'' (QI). 
(More recently, a much simpler proof of Eq.~(\ref{eq:QI}) has been 
given, as well as derivations of similar bounds for the massive scalar 
and electromagnetic fields \cite{FR97}.)
 
Although the QI-bound was initially derived for a massless scalar field in 
Minkowski spacetime (without boundaries), it was argued in Ref.\cite{FRWH} 
that in fact the bound should also hold in a curved spacetime and/or one with 
boundaries, in the limit of short sampling times. More specifically, when the 
sampling time $\tau_0$ is restricted to be smaller than the smallest proper 
radius of curvature or the distance to any boundaries, then the modes of the 
quantum field may be approximated by plane waves, i.e., spacetime is 
approximately Minkowski. In this region, Eq.~(\ref{eq:QI}) should hold. 
Further evidence supporting this conclusion has recently appeared in the 
form of QI-bounds which have been explicitly proven in various static 
{\it curved} spacetimes. In all cases, these bounds reduce to the flat 
spacetime QI's in the short sampling time limit \cite{PFQI,FUN}. 

In Ref.\cite{FRWH} the flat spacetime bound was applied, in the limit 
of short sampling times, to Morris-Thorne traversable wormhole spacetimes. 
The upshot of the analysis was that either the wormhole 
throat could be no larger than a few times the Planck length, or there 
must be large discrepancies in the length scales which characterize the 
wormhole. In the latter case, this typically implied that the exotic matter 
which maintains the wormhole geometry must be concentrated in an 
{\it exceedingly} thin band around the throat. These results would 
appear to make the existence of static traversable wormholes very 
unlikely. A similar analysis using the flat space QI 
has been applied to the Alcubierre ``warp drive'' 
spacetime \cite{PFWD}, which also requires exotic matter. Here as well, 
it was found that the wall of the ``warp bubble'' surrounding a spaceship 
must be unphysically thin compared to the bubble radius. 
In this section, we apply the flat space QI to 
the Krasnikov spacetime, again in the short sampling time limit, and reach a 
similar conclusion regarding the thickness of the negative energy 
regions of the Krasnikov tube. 

Consider a geodesic observer who is initially at rest, i.e., 
$dx/d{\tau}=d{\rho}/d{\tau}=d{\phi}/d{\tau}=0$. These initial conditions 
imply 
\begin{equation}
\frac{d^2 x^{\mu}}{{d\tau}^2} + \Gamma^{\mu}_{\,\,tt} \, 
\biggl (\frac{dt}{d\tau}\biggr )^2 \, =0\,,
\label{eq:geod}
\end{equation}
which reduce to:
\begin{eqnarray}
&{}&\frac{d^2 t}{{d\tau}^2} -  \frac{(1-k)\,{k,_t}}{(1+k)^2}  
\, \biggl (\frac{dt}{d\tau}\biggr )^2 \, =0\,,
\label{eq:geodt}  \\
&{}&\frac{d^2 x}{{d\tau}^2}  - \frac{2\,{k,_t}}{(1+k)^2} \, 
\biggl (\frac{dt}{d\tau}\biggr )^2 \, =0\,,
\label{eq:geodx}  \\
&{}&\frac{d^2 \rho}{{d\tau}^2}  \, =\,\frac{d^2 \phi}{{d\tau}^2} \,=0\,. 
\label{eq:geodrhophi}
\end{eqnarray}
Therefore we see that initially static geodesic observers will remain static 
in the region of the spacetime where 
$k,_{t}\equiv \partial k/{\partial t} = 0$, i.e., 
long after the formation of the tube. In this region, 
from Eq.~(\ref{eq:geodt}), 
$dt/{d\tau} = const$, which we could choose to be $1$ so that 
$dt=d\tau$. By suitably adjusting the zeros of each time coordinate, we can 
make the coordinate time, $t$, equal to the proper time, $\tau$, in this 
region.

To apply the flat spacetime QI, we must first transform to the orthonormal 
frame of the static geodesic observer. The metric, Eq.~(\ref{eq:K4D}), can be 
diagonalized by the transformation 
\begin{eqnarray}
d{\hat t} &=& dt - \biggl(\frac{1-k}{2}\biggr) \, dx \,,  \label{eq:transt} \\
d{\hat x} &=& \biggl(\frac{1+k}{2}\biggr) \, dx \,,   \label{eq:transx} \\
d{\hat \rho} &=& d\rho  \,  \label{eq:transrho} \\
d{\hat \phi} &=& \rho \, d\phi  \,,
\label{eq:transphi}\
\end{eqnarray}
which corresponds to the (non-coordinate) basis \cite{SCHUTZ} 
\begin{eqnarray}
{\bf e}_{\hat t} &=& {\bf e}_t \,, \label{eq:hatet} \\
{\bf e}_{\hat x} &=& \biggl(\frac{1-k}{1+k}\biggr) \, {\bf e}_t \,+
\, \biggl(\frac{2}{1+k}\biggr) \, {\bf e}_x \,, \label{eq:hatex} \\
{\bf e}_{\hat \rho} &=& {\bf e}_{\rho} \,, \label{eq:haterho} \\
{\bf e}_{\hat \phi} &=& \biggl(\frac{1}{\rho} \biggr)\,{\bf e}_{\phi} \,. 
\label{eq:hatephi}
\end{eqnarray}
In this basis, using the fact that 
${\bf e}_{\alpha} \cdot {\bf e}_{\beta} = g_{\alpha  \beta}$, we have 
\begin{equation}
{\bf e}_{\hat \mu} \cdot {\bf e}_{\hat \nu} = 
\eta_{\hat \mu \hat \nu} \,.
\end{equation}
It is also easily seen that, in the region where $k=1$, 
Eqs.~(\ref{eq:hatet}-\ref{eq:hatephi}) reduce to a set of orthonormal basis 
vectors in ordinary Minkowski spacetime. (Note that this basis is also 
well-behaved in the case when $k=0$.)

The stress-tensor and Riemann tensor components in this frame are 
\begin{equation}
T_{\hat\mu \hat\nu} = 
T_{a b} \, e^a_{\hat\mu} \, e^b_{\hat\nu} \,,  \label{eq:That}
\end{equation}
and 
\begin{equation}
R_{\hat\mu \hat\nu \hat\alpha \hat\beta} = 
R_{a b c d} \, e^a_{\hat\mu} \, e^b_{\hat\nu} 
\, e^c_{\hat\alpha} \, e^d_{\hat\beta} \,,
\label{eq:Rhat}
\end{equation}
respectively.  Here  the greek indices label the vector of the basis, and the 
latin indices denote components in the original (coordinate) frame.  From 
Eqs.~(\ref{eq:hatet}) and ~(\ref{eq:That}), we see that the energy density 
in the orthonormal frame is the same as in the coordinate frame, i.e., 
\begin{equation}
T_{\hat t \hat t} = T_{tt}  \,.  \label{eq:Ttthat}
\end{equation}

We will evaluate the energy density in the middle of the left end cap 
at a time long after the formation of the tube, i.e., at 
$x=\epsilon /2 ,\, t \gg  x + \rho + \epsilon$. In this region 
\begin{equation}
\theta_\epsilon(x+\epsilon - D) = 0 \,,
\end{equation}
and 
\begin{equation}
\theta_\epsilon(t-x-\rho) = 1 \,.
\end{equation}
Therefore, in our chosen region, we have 
\begin{equation}
k = 1- (2 - \delta) \, \theta_\epsilon(\rho_{max} - \rho)  \, 
\theta_\epsilon(x) \,.
\label{eq:kevT}
\end{equation}
Let us now choose the following specific form for $\theta_\epsilon (\xi)$ :
\begin{equation}
\theta_\epsilon (\xi)= \frac{1}{2} \,  
\biggl \{ {\rm tanh} 
\biggl [ 2 \, \biggl ( \frac{2 \xi}{\epsilon}  - 1 \biggr ) \biggr ] +1 
\biggr \} \,.
\label{eq:kspec}
\end{equation}
This function has the general desired properties outlined in 
Sec.~\ref{sec:4D} \cite{COMKCOR}. However, we do not expect 
our main conclusion to depend on the detailed form of $k$. 
At $x=\epsilon / 2$, $\theta_\epsilon (x)=0.5$, so 
\begin{eqnarray}
k &=& 1 - \biggl (1- \frac{\delta}{2} \biggr ) \, 
\theta_\epsilon \, (\rho_{max} - \rho ) \\
&=& 1 -  \frac{1}{2} \,\biggl (1- \frac{\delta}{2} \biggr ) \, 
\biggl \{ {\rm tanh} 
\biggl [ 2 \, \biggl ( \frac{2 ( \rho_{max} - \rho )}{\epsilon}  
- 1 \biggr ) \biggr ] +1 \biggr \} \,.
\label{eq:klend}
\end{eqnarray}

Note that from Eq.~(\ref{eq:T}), the energy density depends only 
on derivatives of $k$ {\it with respect to $\rho$}. Consider a static 
observer in the middle of the left end cap at $\rho = \rho_{max} - \epsilon$. 
Let $\rho_{max} = n \epsilon$, where $n > 1$, although $n$ is 
not necessarily assumed to be an integer. 
Substitution of these expressions into Eq.~(\ref{eq:T}) 
gives \cite{COMSING} 
\begin{equation}
T_{\hat t \hat t} = \biggl(\frac{1}{8 \pi}\biggr) \,
\frac{[0.271\,(3.504 - 4.034\,n - 0.008 \, \delta + 0.017\, n \, \delta - 
0.872 \,{\delta}^2 + n\, {\delta}^2)]}{ {\epsilon}^2 \, (n-1) \, 
{(1.018 + 0.491 \,\delta)}^2}  \,.
\label{eq:Tttji}
\end{equation} 
Recall that $0 \lprox \delta \leq 2$. The value $\delta = 2$ corresponds to 
$k=1$ (usual Minkowski spacetime with no opening of the light cone), 
while $\delta \approx 0$ corresponds to $k \approx -1$ in the vacuum 
inside the tube (Minkowski spacetime with maximum ``opening out'' of 
the light cone). Therefore, for effective ``warp travel'', we want $\delta$ 
to be as small as possible. Expansion of Eq.~(\ref{eq:Tttji}) in a power 
series in $\delta$ shows that for small $\delta$ and $n$ large compared 
to $1$, 
\begin{equation}
T_{\hat t \hat t}  \approx -\frac{1}{8 \pi {\epsilon}^2} \,.
\label{eq:Tttsdelta}
\end{equation} 
Let the magnitude of the maximum curvature tensor component 
in the static orthonormal frame be denoted by $\hat R_{max}$. 
Then a (somewhat tedious) calculation using Eq.~(\ref {eq:Rhat}) 
shows that for our chosen observer, in the same limits, 
\begin{equation}
\hat R_{max} \approx \frac{1}{{\epsilon}^2} \,.
\end{equation}
(Note that the curvature tensor components, 
unlike the energy density, will contain derivatives of $k$ with 
respect to $x$.) Hence the smallest proper radius of 
curvature at this location is 
\begin{equation}
r_c  \approx \frac{1}{\sqrt{\hat R_{max}}}  \approx \epsilon \,.
\end{equation}

Let us now apply the QI-bound, Eq.~(\ref{eq:QI}), to the energy density 
seen by our static geodesic observer.  We assume that $T_{\hat t \hat t}$ 
is the expectation value of the stress-tensor operator in some quantum 
state of the quantized massless scalar field \cite{COMEMF}. 
As argued previously in 
Ref.\cite{FRWH}, for this flat spacetime bound to be 
applicable, we must restrict our sampling time to be smaller than the 
smallest local proper radius of curvature, i.e., 
\begin{equation}
\tau_0 = \sigma \, \epsilon \,,
\label{eq:st}
\end{equation}
where $\sigma \ll 1$. In this region, spacetime is approximately flat. 
Note that as long as we consider the region of spacetime 
corresponding to times long after tube formation, the limit of short 
sampling times should also kill off any effects of time-dependence 
of the metric, which occurred during tube formation, on the modes 
of the quantum field. Over the 
timescale $\tau_0$, the energy density is approximately 
constant, so we have 
\begin{equation}
{{\tau_0} \over \pi}\, \int_{-\infty}^{\infty}\,
{{T_{\hat t \hat t}  \, \, d\tau}
\over {{\tau}^2+{\tau_0}^2}}  \approx \, T_{\hat t \hat t} \, \gprox
-{3\over {32 {\pi}^2  {\sigma}^4 {\epsilon^4}}}\,,  
\end{equation}
which implies 
\begin{equation}
\epsilon \lprox \frac{l_P}{{\sigma}^2} \,,
\label{eq:epbound}
\end{equation}
where $l_P$ is the Planck length. For a ``reasonable'' choice of $\sigma$, 
for example $\sigma\approx 0.01$, we have that 
\begin{equation}
\epsilon \lprox 10^4 \, l_P \approx 10^{-31} \, {\rm m} \,.
\label{eq:ebound}
\end{equation}

For an observer in the middle of the 
right end cap, i. e., at $x=D-\epsilon/2$, it is easily shown that the 
expression for $k$ is the same as that given in Eq.~(\ref{eq:klend}). 
Since the energy density depends only on derivatives of $k$ with 
respect to $\rho$, its value will be the same for observers in the 
middle of each end cap, at the same $\rho$-position. For 
times long after tube formation, the spacetime is spatially symmetric 
with respect to reflections of the tube through the plane $x=D/2$. 
Hence the components of the curvature tensor in the static 
orthonormal frame should be the same at $x=\epsilon/2$ and 
$x=D-\epsilon/2$. Therefore our previous argument should apply 
to both end caps.  

At the midpoint of the tube, i. e., at $x=D/2$, $\theta_\epsilon(x)=1$ 
and $\theta_\epsilon(x+\epsilon-D)=0$, so in the static region 
$k=1-(2-\delta)\,\theta_\epsilon(\rho_{max}-\rho)$. One can again 
show that for a static observer at $\rho=\rho_{max}-\epsilon$, 
$T_{\hat t \hat t}  \approx - 1/(8 \pi {\epsilon}^2)$, in the small 
$\delta$ limit. (The nonzero energy density in the region 
just inside the inner wall of the tube is a consequence of 
the ``tails'' of the $\theta_\epsilon$-functions.) By symmetry, 
in this region, $k,_x=0$ at $x=D/2$. It can be shown that the 
curvature tensor components contain no second derivatives 
with respect to $x$. The components can therefore only depend 
on derivatives of $k$ with respect to $\rho$. Again one can 
show that the smallest proper radius of 
curvature at this location is $r_c  \approx \epsilon$. Therefore 
our conclusion, Eq.~(\ref{eq:ebound}), applies to the walls of 
the (hollow) Krasnikov tube as well as to the end caps 
\cite{COMONE}. 

In the preceding discussion, we assumed that 
$\rho_{max} \gg \epsilon$, i. e., that $n$ was large compared to 
$1$. If we relax this requirement and consider thick tubes, 
with $n$ of order $1$, then $\rho_{max} \approx \epsilon$. In 
this case, from dimensional arguments, we should have 
$T_{\hat t \hat t} \approx - 1/ (8 \pi \, {\rho_{max}}^2)$, 
$\hat R_{max} \approx 1/{\rho_{max}}^2$, and 
$r_c \approx \rho_{max}$. Application of our QI  
now yields a bound on the radius of the tube:
\begin{equation}
\rho_{max} \lprox \frac{l_P}{{\sigma}^2} \,.
\end{equation} 
This result is similar to that found in the case of traversable 
wormholes. 

Let us now estimate the total amount of negative energy required 
for the maintenance of a Krasnikov tube \cite{TOTALE}. Our task 
is complicated by the fact that the $t=const$ slices of the Krasnikov 
spacetime are not everywhere spacelike. The metric on such a slice 
is given by 
\begin{equation}
ds^2  =  k(t,x,\rho)\, dx^2 + d{\rho}^2 + {\rho}^2 \, d{\phi}^2 \,,
\label{eq:K3D}
\end{equation}
which can be nonspacelike when $k<0$. Therefore let us instead 
estimate the total negative energy in a thin band in $\rho$ 
over which $k \approx const$. In this band, from 
Eqs.~(\ref{eq:transt}) and ~(\ref{eq:transx}), the metric can 
be written as 
\begin{equation}
ds^2 = -d{\hat t}^2 \, + d{\hat x}^2 \, + d{\rho}^2 \,+ 
{\rho}^2 \, d{\phi}^2 \,.
\label{eq:MTB}
\end{equation}
Consider a band $\Delta \rho$ where $k \approx const$ and the 
energy density is most negative. We see from Fig. 5 that 
such a band has the form 
\begin{equation}
\Delta \rho = \alpha \epsilon \,,
\label{eq:DELTAP}
\end{equation}
where $\alpha \ll 1$. For a small enough choice of $\alpha$, 
we can write the metric in this region in the simple form, 
Eq.~(\ref{eq:MTB}). 
The proper volume of the band is 
\begin{equation}
V \approx 2\pi \rho_{max} \,(\Delta \rho) \, D =
2\pi \alpha \epsilon \, \rho_{max} \, D \,.
\label{eq:V}
\end{equation}
A rough estimate of the total negative energy contained in this band is
\begin{equation}
E \approx T_{\hat t \hat t} \, V \approx 
- \frac{\alpha \,\rho_{max} \, D}{\epsilon} \,.
\label{eq:E1}
\end{equation}
From our QI bound, Eq.~(\ref{eq:epbound}), 
we also have that $\epsilon \approx l_P/{\sigma}^2$, where $\delta$ is 
assumed to be very small. As an example, let 
$D=\rho_{max}= 1\,{\rm m} = 10^{35} \,l_P$, and 
$\epsilon = 100 \, l_P$. Then 
\begin{equation}
E \approx - \alpha \, 10^{68} \,m_{Planck} = -\alpha \, 10^{63} \,{\rm g}
= - \alpha \, 10^{18} \, M_{galaxy} \,,
\end{equation}
where we have taken $M_{galaxy} \approx 10^{12}$ solar masses. 
Thus even if we take $\alpha$ to be very small, say $0.01$, 
one requires negative energies of the order of $10^{16}$ 
galactic masses just to make a Krasnikov tube $1$ meter long and 
$1$ meter wide. For a tube that stretches from here to the 
nearest star, i. e. $D \approx 4 \times 10^{16} {\rm m}$, we need 
$E \approx - 10^{32} M_{galaxy}$. Similar orders of 
magnitude were found in the case of the Alcubierre warp bubble 
\cite{PFWD}. Note that we do not expect the positive and negative 
energies on the outside and inside of the tube to add 
to zero in general, since the cancellation would have to be exact to 
{\it extraordinarily} high accuracy \cite{FHD}, given the large 
magnitudes involved.  

We have been assuming that $\delta \approx 0$, so as to maximize 
the amount by which the light cone is opened out within the tube. 
In particular, values of $\delta < 1$ are needed to allow travel 
backward in time and the possibility of CTCs. The dependence of 
our results on $\delta$ can be easily estimated as follows. Define 
$\eta = 2 - \delta$, so that $k = 1- \eta$ within the (hollow part of the) 
tube, and $k$ changes by $\eta$ across the wall of the tube. For 
$k = 1 - \eta$, the left-hand branch of the light cone in Fig. 2 is given 
by $dx/dt = -1/ (1 - \eta)$. We see that $\partial k/ \partial\rho \sim 
\eta/ \epsilon$ and ${\partial}^2 k/ \partial {\rho}^2 \sim 
\eta/ {\epsilon}^2$ within the tube wall; thus, from Eq.~(\ref{eq:T}), 
in the limit $\eta \ll 1$ and $\epsilon \ll \rho$, 
$T_{tt}$ scales as $\eta/{\epsilon}^2$, and 
$r_c \gg \epsilon$. For small $\eta$, the negative energy 
densities in the walls are thus very small and the QI bound, as well 
as the requirement $\tau_0 \ll r_c$, can be satisfied for macroscopic 
values of $\epsilon$ and $\tau_0$. For example, one can satisfy the 
QI with $\tau_0 = \epsilon \approx 1 \, {\rm cm}$, but only by taking 
$\eta \approx {l_P}^2 \, {\epsilon}^2 / {\tau_0}^4 \approx 10^{-66}$. 
It might therefore actually be possible to establish a region within which 
superluminal travel is, in principle, allowed. However the change 
in the slope of the left branch of the light cone, illustrated in Fig. 2, 
is proportional to $\eta$ for small $\eta$, and hence the speed of a light 
ray directed along the negative $x$-axis within the tube, as seen by 
observers outside, would exceed $1$ by only one part in $10^{66}$. 
The existence of superluminal travel would thus appear to be completely 
unobservable.

\section{Conclusions}
\label{sec:conc}
The Alcubierre ``warp drive'' spacetime suffers from the drawback that a 
spaceship at the center of the warp bubble is causally disconnected from 
the outer wall of the bubble. We have discussed and generalized a 
metric, originally designed by Krasnikov to circumvent this 
problem, which requires that any modification of the spacetime to allow 
superluminal travel necessarily occurs in the causal future of the launch 
point of the spaceship. As a result, this metric has the interesting 
feature that the time for a one-way trip to a distant star is limited by 
all the usual restrictions of special relativity, but the time for a 
{\it round trip} may be made arbitrarily short. In four dimensions this 
entails the creation of a ``tube'' during the outbound flight of the 
spaceship, which connects the Earth and the star. Inside the 
tube, the spacetime is flat but with the light cones ``opened out'' to 
allow superluminal travel in one direction, as seen by 
observers outside the tube. Although the creation 
of a single Krasnikov tube does not entail the formation of 
closed timelike curves, we showed that two spatially separated 
tubes could be used to construct a time machine - a feature shared 
by two-wormhole or two-warp bubble systems. This poses a 
problem for causality even if tubes of only, say, laboratory dimensions 
could be realized in practice. 

In addition, we have also shown that, with relatively modest assumptions, 
maintenance of a such a tube long after formation will necessarily require 
regions of negative energy density which can be {\it no thicker than} a few 
thousand Planck lengths. Estimates of the total negative energy required 
to construct Krasnikov tubes of even modest dimensions were shown 
to be unphysically large. Similar difficulties have been recently shown to 
plague warp bubbles and wormholes \cite{WHCOM}. The Krasnikov 
tube suffers from some of the same drawbacks as these other 
proposed methods of faster-than-light travel, and is hence also a very 
unlikely possibility.

\vskip 0.2 in
\centerline{\bf Acknowledgements}
We would like to thank Michael Pfenning and Larry Ford for helpful 
discussions. TAR would like to thank the members of the Tufts 
Institute of Cosmology for their gracious hospitality while this work was 
being done. This research was supported in part by NSF Grant 
No. PHY-9507351 and by a CCSU/AAUP faculty research grant.


\begin{thebibliography}{--}
\bibitem{COMKCOR} The specific form of $k$ chosen here has the minor 
             drawback that the numerical value of 
             $\partial k/ \partial \rho$ does not quite go exactly 
              to zero at $\rho=0$ (as it must by symmetry), although it does 
              become very small. The resulting apparent ``singularity'' 
              in $T_{tt}$ at $\rho=0$ can be remedied by using 
              a somewhat more cumbersome form of $k$ with an additional 
              term which has the effect of forcing  $\partial k/ \partial 
              \rho= 0$ at $\rho=0$. Since this added complication changes 
              neither our results nor our conclusion, while tending to 
              obscure the physical interpretation of $k$, we have chosen 
              to use the simpler form.
\bibitem{COMSING} The apparent singularity in $T_{\hat t \hat t}$ when 
              $n=1$ is related to the problem at $\rho = 0$ mentioned in the 
              previous footnote. It arises because for $n=1$, we are evaluating 
              $T_{\hat t \hat t}$ at $\rho=0$.
\bibitem{COMEMF} From the results of Ref.\cite{FR97}, our 
            conclusion will be the same if we construct the Krasnikov tube 
            out of quantized massive scalar or electromagnetic fields.
\bibitem{COMONE} We assumed for simplicity that the values of the 
              $\epsilon$'s in all of the $\theta_\epsilon$-functions were 
              equal. Relaxation of this assumption would make the model more 
              complicated but should not, we believe, change the essential 
              result. That is, in the latter case as well, we expect that 
              (although we have not proven it) at least one of the $\epsilon$'s 
              appearing in the expression for $k$ will be constrained by the 
              quantum inequality to be exceedingly small.
\bibitem{TOTALE} Of course the {\it total} energy (matter plus 
              gravitational) of the Krasnikov tube, as measured by observers 
              at infinity, is zero since the external metric is Minkowskian.
\bibitem{WHCOM} The problem of the total amount of negative energy 
               required is not quite as severe in the case of wormholes. For 
               example, for an ``absurdly benign'' wormhole, discussed in 
               Ref. \cite{FRWH}, the energy density is equal 
               to $-1/(4 \pi a_0 r_0)$, where $a_0$ is the thickness 
               of the negative energy band and $r_0$ is the throat radius. 
               The proper volume is approximately $4 \pi a_0 {r_0}^2$; 
               hence the total negative energy required is $E  \approx - r_0$. 
               However, this still requires about one Jupiter mass of 
               negative energy to support a wormhole with a $1$ meter 
               throat radius (see p. 174 of Ref. \cite{VBOOK}). In the cases 
               of the Alcubierre bubble and the Krasnikov tube, the negative 
               energy density scales as $1/ {\epsilon}^2$, so the total 
               negative energy required scales as $1/ {\epsilon}$, where 
               $\epsilon$ is the thickness of the wall.
\end{thebibliography}
\end{document}